\documentclass[a4paper]{jacow}

\ifboolexpr{bool{xetex} or bool{luatex}}
 {}
 {\usepackage[utf8]{inputenc}}

\usepackage{subfig}
\usepackage{amsmath}
\usepackage{graphics}
\usepackage{amssymb}
 \usepackage{graphicx}
\usepackage{marginnote}
\usepackage{float}
\usepackage{lipsum}

\ifboolexpr{bool{jacowbiblatex}}
 {%
  \addbibresource{jacow-test.bib}
  \addbibresource{biblatex-examples.bib}
 }{}

\copyrightspace[2cm] 

\begin{document}

\title{LOW ENERGY BEAM TRACKING UNDER SCATTERING FOR A COLD ELECTRON SOURCE IN MANCHESTER}

\author{R.B. Appleby\thanks{Corresponding author, Robert.Appleby@manchester.ac.uk. $^{\dagger}$The Cockcroft Institute of Accelerator Science and Technology. $^\S$ European Organisation for Nuclear Research, CERN. $^\P$ The University of Manchester - Photon Science Institute.}$^{\dagger}$, W. Bertsche$^\S$, M. Harvey$^\P$, M. Jones, B. Kyle, O. Mete$^{\dagger}$, A. Murray$^\P$, G. Xia$^{\dagger}$, \\ The University of Manchester, Manchester, M13 9PL, United Kingdom} 

\maketitle

\begin{abstract}
High quality electron beams, with high spatial and temporal resolution, have an important use in electron diffraction experiments to probe and study the constituents of matter.  A cold electron source is being developed based on electron ionisation from an atom cloud trapped by using AC magneto-optical methods in the University of Manchester. The technique will produce bunches of electrons
well suited for high precision and single shot electron diffraction. In this paper issues of modelling at low energies for this state of art electron source with very low energy spread are presented, with a focus
on newly developed tools to model the scattering in the meshes used to support the extraction electric fields. The dependence on emittance growth on mesh wire thickness is studied. 
\end{abstract}

\section{Introduction} 
Electron diffraction experiments are an integral part of many fields of research, including crystallography, spectroscopy and investigations into chemical bonding. As these fields progress, there is an ever-increasing need for electron beams with better spatial and temporal resolution, requiring investigation into novel methods of increasing beam quality. The AC-MOT, currently being developed at Manchester, promises to deliver low-emittance 
and low temperature beams through the cooling of the electron source using magneto-optical trapping~\cite{jinst,ipac2014}. It offers advantages over conventional magneto-optical trapping techniques through the AC magnetic field, such that there are no residual fields due to eddy currents persisting after the trapping cycle has ended. As such, the trajectories of electrons extracted from the AC-MOT are unaffected by stray external fields resulting in a more reliable electron source with potentially higher beam quality. We present in this paper newly developed tools to study particle transport and scattering in the Manchester-Cockcroft Institute experiment and show results of simulations conducted to model the extraction process from the AC-MOT. In this work several codes have been employed, namely CST~\cite{CST}, for modelling the experimental region and calculating
the electric fields used for electron extraction; General Particle Tracer~\cite{GPT}, for simulation of particle trajectories in drift spaces; and Geant4~\cite{geant4_1,geant4_2}, for determining emittance growth due to scattering. Note that some preliminary calculations have been performed with FLUKA~\cite{fluka} but are not described in this paper.
We outline an assessment of the extraction configuration which is being built and make the first look at the impact of mesh scattering on the transported electrons. 

\section{Extraction region layout}

\begin{figure}[htb!] 
\centering
\includegraphics[width=0.4\textwidth] {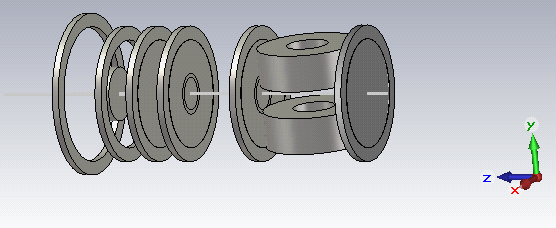}
\caption{The layout of the extraction region, showing the extraction electrodes.}
\label{fig:layout}
\vspace{-1.5em}
\end{figure}

The atoms in the MOT are ionised and the electrons are extracted using a series of electrostatic electrodes towards the diagnostic section (or later an electron diffraction experiment or injection into a FEL). A full description of the AC-MOT in this experiment can be found in~\cite{jinst}. 
After passing through the grounded MOT field coils, the electrons pass through three electrodes connected to 5 kV power supplies, which provide the extraction fields. An electrode is constructed of a stainless steel ring covered with a fine mesh, designed to support the fields but allow electron transmission. The three electrodes are followed by the Microchannel Plate (MCP) detector, with its front plate biased at 200 V. 
The electrode arrangement is shown in Fig. \ref{fig:layout}.
%

\begin{figure}[htb!] 
\centering
\includegraphics[width=0.4\textwidth] {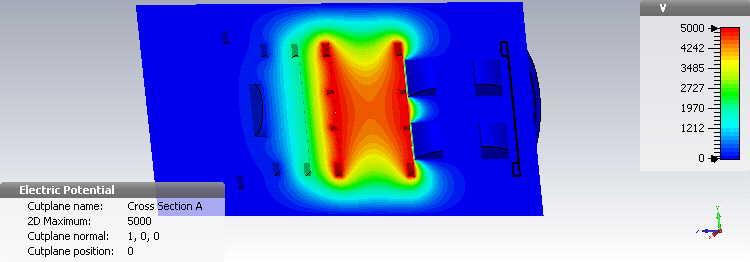}
\caption{Voltage configuration across the plates.}
\label{fig:field}
\vspace{-1.5em}
\end{figure}

The need to accelerate and extract the electrons before the bunch is rapidly expanded by space-charge forces means the extraction voltages need to be high. The MOT coils are grounded and the voltage plates are positioned to provide a strong electric field over the MOT region and subsequent transport of the electrons towards the MCP is performed in a drift region delimited by the first and second electrodes, both held at 5 kV. A third plate very close to the second plate is held at 2 kV and provides a retarding electric field to slow the electrons down in a short time before the MCP to provide optimum MCP efficiency. The fast extraction configuration gives a strong field across the MOT, resulting in a larger energy spread than the penetrating field configuration described in \cite{ipac2014}. Figure \ref{fig:field} shows the voltages applied to the plates and the resulting potential, computed using the electrostatic modelling tools of CST-Microwave Studio \cite{CST}. 

\section{Particle Tracking}

Particle tracking through the extraction region was performed by using General Particle Tracer (GPT) \cite{GPT} with 3D field maps extracted from CST. The choice of tracking tool is dictated by the need to model correctly low energy transport through electromagnetic fields and an efficient space-charge model to correctly include the intra-bunch forces in the ultra-low-energy beam. GPT is used
for  this purpose. A focus of the current design work presented in this paper is the impact of low energy electron scattering in the meshes used to support the electric fields - in this work the scattering model in GPT is replaced by a more accurate scattering model of GEANT4, to give a realistic estimation of the emittance growth from
elastic scattering in the meshes. So in this paper, the effect of the beam-mesh interaction was investigated with a coupled tracking tool of both GPT  and GEANT4 \cite{geant4_1,geant4_2}, with the fields produced in CST, ensuring the electron-mesh interaction is performed according to the interaction differential cross sections.  
Practically the code coupled is realised with a series of helper scripts, handling coordinate and unit translation at every mesh encountered by the GPT tracking. 

\subsection{GEANT4 Results}

\begin{figure}[htb!] 
\centering
\includegraphics[width=0.3\textwidth] {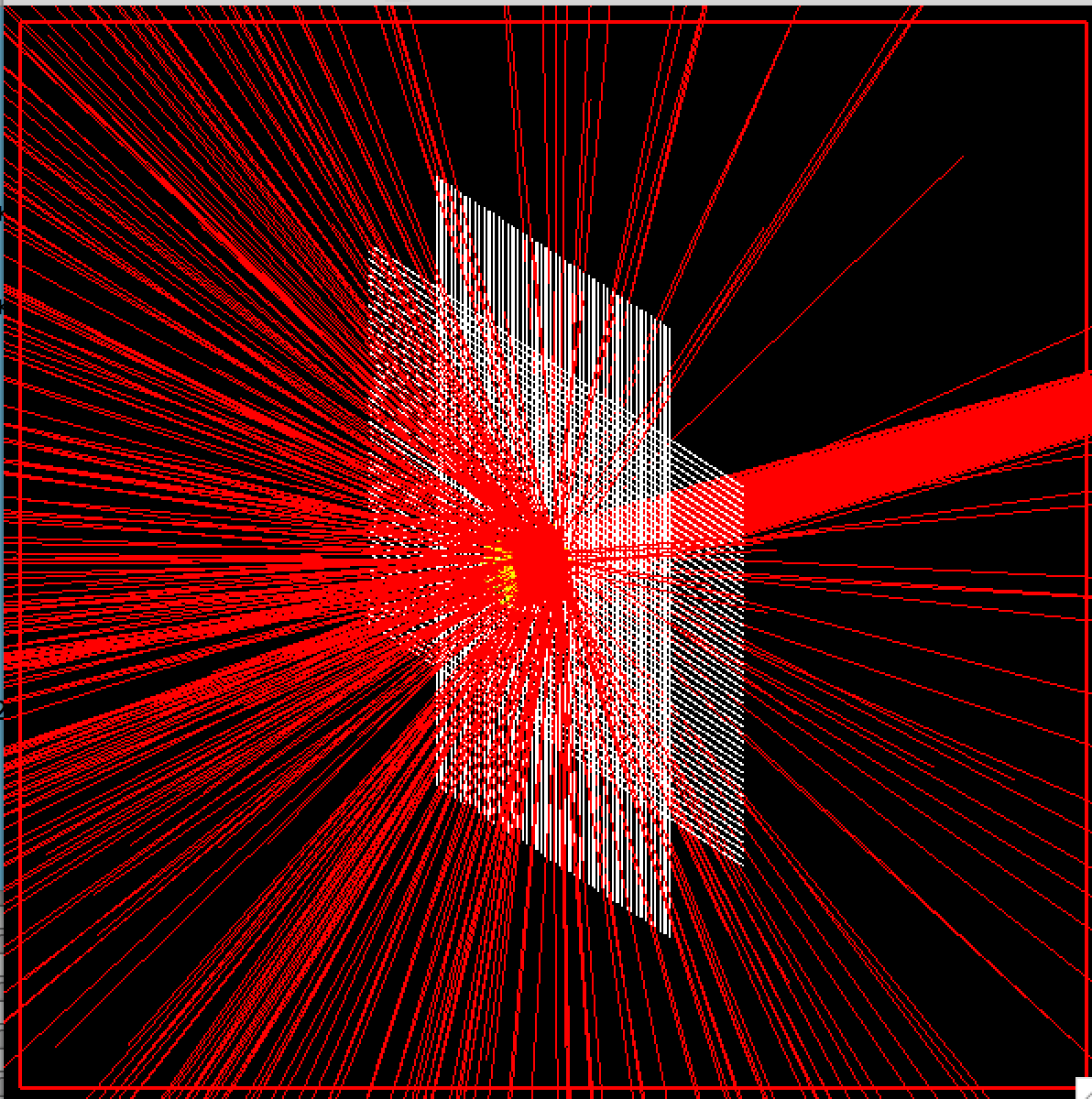}
\caption{An example to particles scattering through the woven mesh with a wire diameter of $25\,\mu$m simulated with GEANT4.}
\label{fig:geant_scatt}
\vspace{-1.5em}
\end{figure}

In this section we show the impact of the scattering through a single mesh using GEANT4. The mesh is modelled as a series of tungsten wires of thickness 0.025 mm with a wire spacing of 0.483 mm, to give a wire woven mesh. 
Figure \ref{fig:geant_scatt} shows graphically the effect of passing through this mesh on a pencil beam of particles with $0.5\,$mm radius with zero angular spread at $5\,$keV energy, where the electrons undergo elastic scattering. The figure shows the electrons which hit the mesh wires can perform large angle scattering. 

\begin{figure}[htb!] 
\centering
\includegraphics[width=0.35\textwidth] {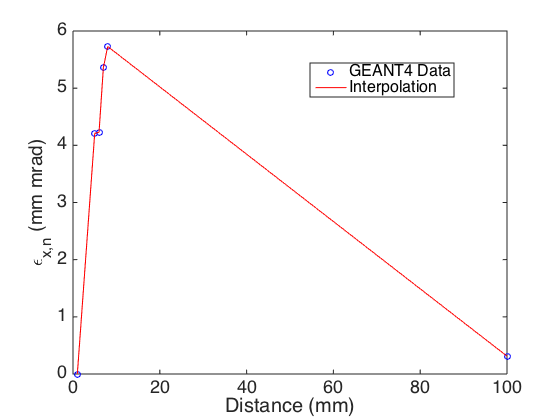}
\caption{Reconstructed rms transverse emittance along the beam axis produced from GEANT4 tracks.}
\label{fig:emitt_geant}
\vspace{-1.5em}
\end{figure}

The rms transverse geometric emittance evolution of these particles that is generated by from GEANT4 tracks is shown in Fig. \ref{fig:emitt_geant} as a function of longitudinal distance. The mesh is located at 5 mm. The beam emittance
growth is seen after the mesh, and the emittance subsequently falls as large amplitude electrons which performed large angle scattering are lost from the simulation. In total  $10\%$ of the initial electrons are lost during the transport from the atom cloud at $z=0$ to the MCP. 

\subsection{GPT+GEANT4 Results}

In this section we show results from transporting the AC-MOT beam through the extraction region with GPT and using GEANT4 for the mesh scattering, giving a stepwise coupled tracking/scattering model. A distribution of 5000 particles was initialised in the GPT input file, with an initial particle distribution of Gaussian both longitudinally, with a 1$\sigma$ bunch length of $0.2\,$mm, and transversely, with a 1$\sigma$ radius of $2\,mm$. This reflects the size of the AC-MOT in the current experiment.  
The initial transverse velocity phase-space distribution ($\gamma\beta_{x,y}$) was also defined as Gaussian, with a width of $1\times10^-6$. The transverse normalised emittance was then scaled to give $2\times10^{-2}$mm$\,$mrad. The initial particle kinetic energy was defined to be $1\,$meV.

This initial particle distribution is then propagated in GPT to a position just before the first electrode. At this point, the GPT output was analysed to determine the parameters of the beam distribution at this position (i.e. the mean, and standard deviation of the Gaussian distribution). These parameters, along with beam energy, were then used to define a new particle distribution in GEANT4, which then modelled the transport of electrons through the tungsten mesh electrodes. Parameters describing phase-space distributions were obtained after the scattering, together with transverse beam emittances. GPT was then used to track this scattered distribution to the next electrode, under the influence of the extraction fields and space-charge. This process was the repeated for the entire extraction region.

\begin{figure}[htb!] 
\centering
\includegraphics[width=0.35\textwidth] {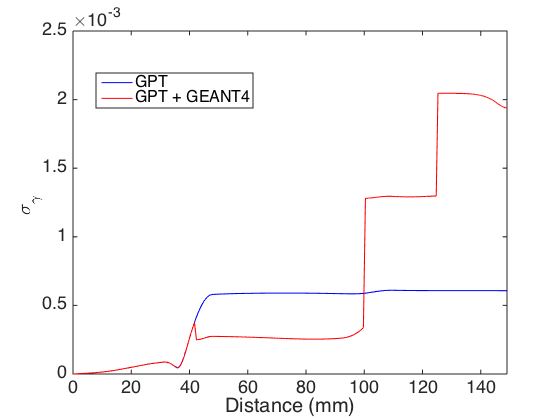}
\caption{Reconstructed rms energy spread along the beam axis using GEANT4 and GPT combined tracking.}
\label{fig:energy_spread}
\vspace{-1.5em}
\end{figure}
\begin{figure}[htb!] 
\centering
\includegraphics[width=0.35\textwidth] {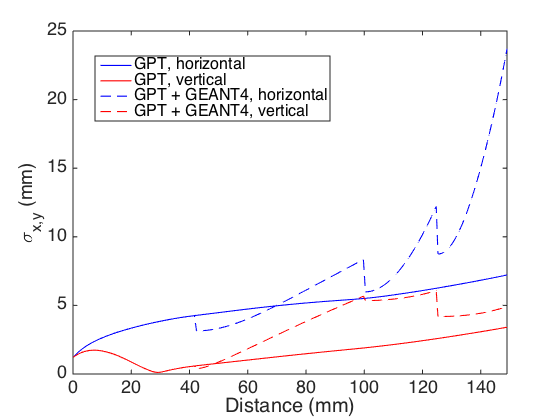}
\caption{Reconstructed rms horizontal and vertical beam sizes along the beam axis using GEANT4 and GPT combined tracking.}
\label{fig:beam_sizes}
\vspace{-1.5em}
\end{figure}

Figures \ref{fig:energy_spread} and \ref{fig:beam_sizes} show the tracking results from the combined tracking of GPT and GEANT4 (with scattering) and also from only-GPT tracking where all particles
pass through the mesh without scattering. The energy spread without scattering shows an increase at the first electrode due to the strong electric field and stays constant until the MCP. The addition of scattering results in increases at each of the extraction electrodes. Similar behaviour is seen in the beam size with and without scattering, where the beam size (due to the mesh interaction) increases in step at each of electrodes when GEANT4 is used to model the mesh scattering. The result is a significantly larger beam with a larger energy spread at the MCP when the mesh scattering is included. 

\begin{figure}[htb!] 
\centering
\includegraphics[width=0.35\textwidth] {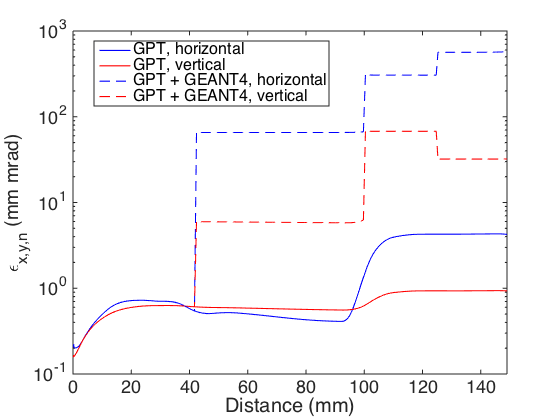}
\caption{Reconstructed rms transverse emittance along the beam axis using GEANT4 and GPT combined tracking.}
\label{fig:combined_emittance}
\vspace{-1.0em}
\end{figure}

The corresponding normalised emittance growth is shown in Figure~\ref{fig:combined_emittance}, again with GPT only and GPT combined with GEANT4. For the case of no scattering, the emittance increases by a factor of around 10 in the transport from the AC-MOT to the MCP, primarily due to space-charge within the bunch. 
However, a much larger emittance growth is seen when the mesh scattering is included in the existing mesh design. 

\begin{figure}[htb!] 
\centering
\includegraphics[width=0.35\textwidth] {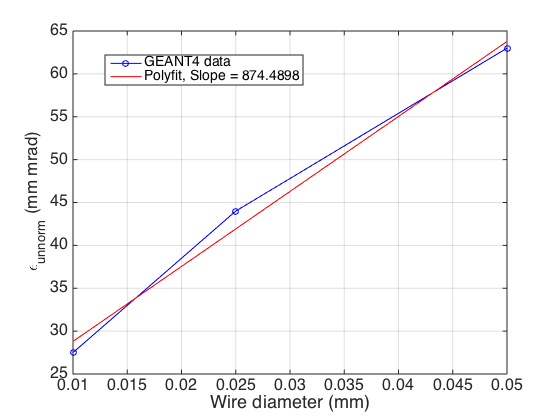}
\caption{The physical emittance growth as a function of mesh wire size, obtained from
GEANT4 simulations.}
\label{fig:wirescan}
\vspace{-1.5em}
\end{figure}

Figure~\ref{fig:wirescan} shows the growth in (unnormalised) emittance for the passage of a pencil beam
through an electrode mesh of varying wire thickness. The existing wires in the Manchester-Cockcroft configuration are 0.025mm, denoted by the blue spot. The expected decrease in the emittance growth for thinner wires is seen, suggesting the emittance growth could be controlled with alternative wire geometry and thickness. Such studies are underway at the present time.

\section{Conclusions}
A technique combining the GEANT4 and GPT codes in order to simulate the non relativistic electrons undergoing acceleration under static electric fields in the Manchester-Cockcroft AC-MPT and scattering through fine mesh structures is presented. The aim is electron transport from the Manchester-Cockcroft AC-MOT to an MCP with controlled emittance growth. We show that the emittance growth of a realistic beam is very high with the existing experimental setup and propose a solution to preserve the emittance through extraction geometry optimisation.  We are also commissioning the experiment in Manchester, with one of the goals to benchmark these simulations. 

%
\raggedend

\end{document}